\documentstyle[12pt]{article} 
\input{epsf}

\newcommand{\cket}[1]{| #1 \rangle}
\newcommand{\bra}[1]{\langle #1 |}

\newcommand{\vev}[1]{\langle #1 \rangle}
\newcommand{\GeV}{\: {\rm GeV}}
\begin{document}

%\begin{center}
%{\large QCD Sum Rule for ${1 \over 2}^-$ Baryons \\}
%\vspace*{0.5cm}
%{ D. Jido\footnote{e-mail address: jido@th.phys.titech.ac.jp} \  and M. Oka \\
%Department of Physics, Tokyo Institute of Technology\\ 
%Meguro, Tokyo 152  Japan} 
%\end{center} 

\title{\Large QCD Sum Rule for ${1 \over 2}^-$ Baryons }
\author{ D. Jido\thanks{e-mail address: 
jido@th.phys.titech.ac.jp}\
and M. Oka \\
Department of Physics, \\ Tokyo Institute of Technology\\ 
Meguro, Tokyo 152  Japan} 
\date{}

\maketitle

\abstract{The masses of the flavor octet and singlet 
baryons with negative parity and spin $\frac{1}{2}$ are calculated 
using the QCD sum rule.  
%We find that the value 
%of the quark condensate determines the mass splitting 
%of the positive and negative parity baryons.  
We find that the chiral symmetry breaking vacuum condensates
cause the mass splitting of the positive and negative baryons.
The present sum rule reproduces the observed masses of these baryons
within 10\%, and predicts the mass of the excited $\Xi$ 
baryon with $J^{P} = {\frac{1}{2}}^{-}$ at 1.63 GeV. We confirm that
the negative-parity state, $\Lambda_S-$, is the ground 
state in the flavor singlet baryon spectrum.  }

\section{Introduction} 
The negative parity excited states of baryons have been understood 
fairly well in the (nonrelativistic) quark model as one-quark excited states 
belonging to the SU(6) $\underline{70}$ representation~\cite{ik}. 
The observed spectrum tends to agree with the prediction, 
although some of the states, such as $\Lambda(1405)$, $N(1535)$, 
have irregular masses and non-natural 
decay rates. Many refinements were proposed to achieve a 
quantitative agreement.

Yet, our understanding of the hadron physics insists that the role of 
the chiral symmetry must be important even in the baryon states. 
Indeed, the chiral symmetry suggests that the positive and negative 
parity states are paired into a parity doublet and the pair would be 
degenerate when the chiral symmetry is restored. Because the 
nonrelativistic quark model does not observe the chiral symmetry, 
a different approach is anticipated to understand the 
chiral structure of baryons, both the positive and negative parity 
states.

%The non-relativistic quark model is successful in the baryon 
%spectrum\cite{ik}.  In the quark model we pay our attention to the 
%symmetries, spin, flavor and the orbital wave function, and baryons 
%are composed by three constituent quarks whose mass is a few hundred 
%MeV. We understand that a negative parity is given to a baryon by the 
%orbital excitation of one quark in the baryon from s-wave to p-wave in 
%the quark model.  The chiral symmetry, however, dose not exist in the 
%quark model, since the constituent quark is given the effect of the 
%spontaneous chiral symmetry breaking.  The chiral symmetry breaking 
%would relate to that the parity doublet is not seen in the hadron 
%spectrum, because of the fact that the nature of the Nambu-Goldstone 
%boson of the spontaneous chiral symmetry breaking gives the lighter 
%mass to the $\pi$ meson than the $\sigma$ meson, which would be chiral 
%partner of the $\pi$ meson.  Therefore it is significant for clarifying 
%the relation of the chiral symmetry to the hadron spectrum to 
%understand the negative-parity baryon spectrum in the view of QCD.

The technique of the QCD sum rule relates the hadron properties to the 
QCD parameters and is a powerful tool to extract the hadron properties 
from QCD, the first principle of the strong interaction\cite{svz,rry}.  
The QCD sum rule for the baryon was first proposed by Ioffe~\cite{i}.
In the QCD sum rule a correlation function, such as $\Pi(p) =i \int 
d^{4}x e^{ip\cdot x} \bra{0} {\rm T} J(x) \bar{J}(0) \cket{0}$, is 
calculated, where $J(x)$ is an interpolating field (IF) that couples 
to the state of the hadron in question.  The nonperturbative effects, 
such as the quark condensate $\vev{\bar{q}q}$ and $\vev{{\alpha_{s} 
\over \pi} GG}$, are included as the power corrections in the 
theoretical side.  The quark condensate $\vev{\bar{q}q}$, which is the 
order parameter of the chiral symmetry breaking, gives effects of the 
chiral symmetry breaking to the hadron spectrum in the QCD sum rule.

In our previous paper\cite{jko}, 
the technique to extract masses of negative-parity baryon in the QCD 
sum rule was proposed and the masses of the nucleons with positive and 
negative parity are calculated.  It is important to separate 
contribution of the negative-parity baryon ($B_{-}$) from that of 
the positive-parity baryon ($B_{+}$), since the IF for baryon couples 
to the states of the $B_{-}$~\cite{i,cdks}, although we define the 
IF $J_{B}$ has positive parity.  In order to separate the $B_{-}$ 
contribution we use the ``old-fashioned'' correlation function defined 
as
\begin{equation}
   \Pi(p) = i \int d^{4} x e^{i p \cdot x} \theta(x_{0}) \bra{0} J_{B}(x) 
   \bar{J}_{B}(0) \cket{0}, \label{eq:oldfco}
\end{equation}
and construct sum rules in the complex $p_{0}$-space in the rest 
frame ($\vec{p}=0$).  Our approach is suitable for investigating
the mass splitting, because $B_{+}$ and $B_{-}$ can be treated 
simultaneously in this sum rule.
In the present paper, we extend the technique to 
the hyperons and the flavor singlet baryons 
$\Lambda_{S}$. 

In Sec.2 the formulation of our approach is explained and we see
that the chiral symmetry breaking is responsible for the 
mass splitting of $B_+$ and $B_-$.
In Sec.3 we present the method to extract the masses and to
determine the parameters in the theoretical side.
In Sec.4 we give the numerical results for the masses of the baryons and
the parameters in the theoretical side.  We also discuss the relation of
the mass splitting to the chiral symmetry breaking vacuum condensate.
We show that the negative parity baryon is the lowest state in the
flavor singlet spectrum and that the QCD sum rule predicts that
the spin-parity of $\Xi(1690)$ is ${1 \over 2}^{-}$.
A summary is given in Sec.5.

\section{Formulation}
It is important to choose an appropriate interpolating field (IF)
in the correlation function in the QCD sum rule.  The IF
should have the same quantum numbers as the baryon in question, so
that it creates or annihilates a single particle state of the
baryon from the vacuum. For the spin ${1 \over 2}$ 
octet baryon two independent IFs can be constructed
without a derivative~\cite{ept}. The IF for the nucleon is, for 
instance, written as
\begin{equation}
   J_{N}(x) = \varepsilon_{abc} [(u_{a}(x)Cd_{b}(x)) 
      \gamma_{5} u_{c}(x) + t (u_{a}(x) C \gamma_{5} d_{b}(x)) 
u_{c}(x)],
  \label{eq:nucur} 
\end{equation}
where $a$, $b$ and $c$ are color indices, $C = i \gamma_{2} 
\gamma_{0}$ (standard notation) is for the charge conjugation and $t$ 
is a real parameter representing the mixing of two independent IFs.  
If we choose $t=-1$ and use the Fierz transformation, the IF 
(~\ref{eq:nucur}) is reduced to the Ioffe's IF~\cite{i}.  In the 
previous paper~\cite{jko}, we found that $J_{N}$ with $t=0.8$ is 
appropriate for the nucleon resonance.  For the $\Sigma$ baryon, we 
replace a d-quark by an s-quark in eq.(~\ref{eq:nucur}) and obtain
\begin{equation}
   J_{\Sigma^{+}}(x) = \varepsilon_{abc} [(u_{a}(x)Cs_{b}(x)) 
      \gamma_{5} u_{c}(x) + t (u_{a}(x) C \gamma_{5} s_{b}(x)) 
u_{c}(x)].
  \label{eq:sicur} 
\end{equation}
Similarly for the $\Xi$ baryon, replacing u-quark by s-quark in 
eq.(~\ref{eq:nucur}), we obtain
\begin{equation}
   J_{\Xi^{-}}(x) = \varepsilon_{abc} [(s_{a}(x)Cd_{b}(x)) 
      \gamma_{5} s_{c}(x) + t (s_{a}(x) C \gamma_{5} d_{b}(x)) 
s_{c}(x)].
  \label{eq:xicur} 
\end{equation}
The IF for $\Lambda$ is more complicated:
\begin{eqnarray}
  J_{\Lambda}(x) & = & \varepsilon_{abc} [(d_{a}(x)Cs_{b}(x)) 
      \gamma_{5} u_{c}(x) + (s_{a}(x)Cu_{b}(x)) \gamma_{5} d_{c}(x)  
      \nonumber \\
  & &   - 2 (u_{a}(x)Cd_{b}(x)) \gamma_{5} s_{c}(x) 
\label{eq:lacur} \\
  & & + t \{ (d_{a}(x)C\gamma_{5} s_{b}(x)) u_{c}(x) + 
       (s_{a}(x)C\gamma_{5} u_{b}(x)) d_{c}(x)  \nonumber \\
  & &  - 2 (u_{a}(x)C\gamma_{5}d_{b}(x)) s_{c}(x) \}] \nonumber  
\end{eqnarray}
The IF for the flavor singlet baryon $\Lambda_S$ is given by the flavor 
antisymmetric combination of the quark operators:
\begin{eqnarray}
  J_{\Lambda_S}(x) & = & \varepsilon_{abc} 
	[(u_{a}(x)C\gamma_{5}d_{b}(x))s_{c}(x)
     - (u_{a}(x) C d_{b}(x)) \gamma_{5} s_{c}(x) \nonumber \\
  & &  - (u_{a}(x)C\gamma_{5}\gamma_{\mu} d_{b}(x)) \gamma^{\mu} s_{c}(x) ] .
  \label{eq:singcur} 
\end{eqnarray}
This IF is unique and has no parameter such as $t$ in the octet IF.

We now explain the technique of the treating the $B_{-}$ in the QCD 
sum rule according to the Ref.~\cite{jko}.  First, we observe that the 
IFs given in eqs (~\ref{eq:nucur})--(~\ref{eq:singcur}) annihilate not 
only the positive parity baryon state, but also a single particle 
state of the negative parity baryon~\cite{i,cdks}.  Since the parity of 
the fermion state can be reversed by multiplying $i \gamma_{5}$, the IF with 
negative parity may be obtained as
\begin{equation}
   J_{-}(x) \equiv i \gamma_{5} J(x). \label{eq:intnp}
\end{equation}
Then we expect
\begin{equation}
%    \bra{0} J(x) \cket{B_{+}} & = & \lambda_{+} u_{+}(x), 
% \label{eq:pp} \\
   \bra{0} J_{-}(x) \cket{B_{-}} =  \lambda_{-} u_{-}(x), 
\label{eq:nn}
\end{equation}
where $\cket{B_{-}}$ denotes a single particle state of the
negative parity baryon, $\lambda_{-}$ is the coupling strength and 
$u_{-}(x)$ is the (corresponding) Dirac spinor.
From eqs.(~\ref{eq:intnp}) and (~\ref{eq:nn}),
we obtain
\begin{equation}
   \bra{0} J(x) \cket{B_{-}} = -\bra{0} i \gamma_{5} J_{-} 
\cket{B_{-}}
   = -i \lambda_{-} \gamma_{5} u_{-}(x).
\end{equation}
Thus, $J(x)$ couples also to the negative-parity baryon.
This short exercise tells us that the conventional sum rule for the 
ground state baryon already contains the negative parity baryon state 
as a part of the continuum spectrum. 

Our task is to separate the 
negative parity contribution properly.
% Therefore the contribution of the $B_{-}$ contribution is already 
% included in the QCD sum rule we have usually used. The reason why
% the $B_{-}$ contribution is not seen is that the
% contribution is regarded as higher resonance and is removed as a 
% continuum. 
%To extract the properties of $B_{-}$ in the QCD sum rule, we must 
%separate the $B_{-}$ contribution from that of the ground state 
%baryon.  
According to the previous paper~\cite{jko}, we use the 
``old-fashioned'' correlation function (~\ref{eq:oldfco}).  In the 
phenomenological side, if the lowest energy states of $B_{+}$ and 
$B_{-}$ are picked up and the rest is regarded as a continuum, the 
imaginary part of the ``old-fashioned'' correlation function is 
written as
\begin{eqnarray}
  {\rm Im} \, \Pi(p_{0})  & = & 
    (\lambda_{+})^{2} {\gamma_{0} + 1\over 2} \delta(p_{0} - 
m_{+}) + 
    (\lambda_{-})^{2} {\gamma_{0} - 1\over 2} \delta(p_{0} - 
m_{-}) \nonumber \\
  & &  + \cdots \: {\rm (continuum)} \label{eq:imphe} \\
  & \equiv & \gamma_{0} A(p_{0}) + B(p_{0}) \label{eq:aandb}.
\end{eqnarray}
%%%%%%%%%%%%%%%%%%%% Added for revised version %%%%%%%%%%%%%%%%%%%%%
In this expression, the zero-width pole approximation is applied.
Although the resonances have significant widths 
(eg.\  about 150 MeV for $N$(1535)),
we expect that the approximation is valid because the IFs for baryons are
composed of three quarks and may not couple strongly with $q^4 \bar{q}$
states.
%%%%%%%%%%%%%%%%%%%%%%%%%%%%%%%%%%%%%%%%%%%%%%%%%%%%%%%%%%%%%%%%%%%%

The difference of the contributions of $B_{+}$ and $B_{-}$ 
in eq (~\ref{eq:aandb})
is the sign 
of the chiral odd part. Therefore $A(p_{0}) + B(p_{0})$ has only 
$B_{+}$ contribution and $A(p_{0})-B(p_{0})$ includes only $B_{-}$ 
contribution. In this way we can separate the $B_{-}$ contribution 
from the ground state baryon.
Moreover we construct the sum rule in the $p_{0}$ complex plane,
because the function $A(p_{0})$ defined in eq.(~\ref{eq:aandb})
is not an analytic function of $p^{2}$,
while the functions $A(p_{0})$ and $B(p_{0})$ is analytic in the 
upper half $p_{0}$-space. 

We take the lowest mass pole and approximate others as a continuum whose 
behavior above a threshold $s_{0}^{\pm}$ is same as the theoretical 
side.  Then we obtain sum rules for the positive- and negative-parity 
baryons:
\begin{eqnarray}
     \int_{0}^{s_{+}} [A^{\rm OPE}(p_{0}) + B^{\rm OPE}(p_{0})] 
       \exp\left[-{p_{0}^{\;2 } \over M^2} \right] dp_{0} & = & 
     (\lambda_{+})^{2} \exp\left[{- \frac{m_{+}^{\;2}}{M^{2}}}\right],  
\label{eq:srp} \\
     \int_{0}^{s_{-}} [A^{\rm OPE}(p_{0}) - B^{\rm OPE}(p_{0})] 
       \exp\left[-{p_{0}^{\;2 } \over M^2} \right] dp_{0} & = & 
     (\lambda_{-})^{2} \exp\left[{- \frac{m_{-}^{\;2}}{M^{2}}}\right], 
\label{eq:srn}
\end{eqnarray} 
where $M$ is the Borel mass and $A^{\rm OPE}$ and $B^{\rm OPE}$ are 
calculated based on QCD using the OPE at the region where $p_0$ is 
large.  It should be noted that the OPE is valid for large $p_{0}$ 
even if $\vec{p}=0$ because the singularity of the correlation 
function resides at the light cone~\cite{hkl}.  The explicit forms 
of $A^{\rm OPE}$ and $B^{\rm OPE}$ are given in appendix for the 
flavor octet and the singlet baryons up to dimension six, ${\cal 
O}(m_{s})$ and ${\cal O}(\alpha_{s}^{\; 0})$.  
%%%%%%%%%%%%%  Added for revised varsion %%%%%%%%%%%%%%%%%%%%%%
The odd higher dimensional terms in the OPE  
do not contribute to the sum rule after the Borel transform,
since their imaginary parts are proportional to 
the odd rank derivatives of the delta-function with respect to
$p_0$. It is easy to show that the integral vanishes after partial
integration.
%%%%%%%%%%%%%%%%%%%%%%%%%%%%%%%%%%%%%%%%%%%%%%%%%%%%%%%%%%%%%%%%%%%

Note that the difference of the sum rules for $B_+$ and $B_-$ is the 
chiral odd term $B(p_{0})$, which is proportional either to the quark 
condensate $\vev{\bar{q}q}$ or to the mixed condensate $\vev{ \bar{q}g 
\sigma \cdot G q}$.  If the chiral symmetry is restored at high 
temperature, for instance, the $B(p_{0})$ term goes to zero in the 
chiral limit.  Then the sum rules (~\ref{eq:srp}) and (~\ref{eq:srn}) 
are identical, and will predict the same masses for the positive and 
negative parity baryons.  This situation is similar to that in the 
linear sigma model for parity-doublet baryons proposed by DeTar and 
Kunihiro~\cite{dk}.  There the positive and negative parity nucleons 
are assumed to form a parity doublet and the Lagrangian has a chiral 
invariant mass term.  Under the restoration of the chiral symmetry, 
the nucleons have the same mass, while in the spontaneous symmetry 
broken phase the mass splitting is proportional to the nonvanishing 
vacuum expectation value of sigma.  In another article~\cite{joh}, we 
have shown that this similarity to the linear sigma model is 
confirmed also in the $\pi NN^*$ coupling in the QCD sum rule.

We note several other approaches of the QCD sum rule for $B_-$.  In 
the case that $B_-$ is the lowest energy state in the considering 
spectrum, the mass of $B_-$ is extracted in the usual sum rule.  In 
Ref.~\cite{i}, Ioffe pointed out that the negative-parity resonance is 
the ground state in the spectrum of the baryon with spin $J={3 \over 
2}$ and isospin $T={1 \over 2}$ and calculated its mass.  Liu also 
applied the QCD sum rule to $\Lambda (1405)$ and concluded 
that for the $\Lambda (1405)$ the IF consisting of three quarks and a 
flavor octet quark-antiquark pair is important~\cite{liu}.  In his analysis, 
however, the continuum term is not considered.  We obtain the 
realistic mass of $\Lambda(1405)$ in the three-quark sum rule with a 
continuum term.
%In Ref.~\cite{cdks} the negative-parity nucleon was 
%treated with making the chiral odd sum rule negative.  
Some other approaches treat $B_-$ with an IF which does not couple to 
the positive-parity ground state baryon.  In Ref.~\cite{cdks} an 
optimized IF for $N_{-}$ was proposed by requiring that the chiral odd 
correlation function, in which the $B_+$ contribution and the $B_-$ 
contribution have different sign, becomes negative.  Lee and Kim also 
investigated the mass of N(1535)~\cite{leekim} and 
$\Lambda(1405)$~\cite{kimlee} in the QCD sum rule.  They proposed a 
new IF with a covariant derivative, expecting that it has a large 
overlap with the nonrelativistic quark wave function of N(1535).  They 
chose the IF so that it does not couple to the ground state 
nucleon~\cite{leekim}.  We, however, employ the nonderivative IF in 
the present study because our main interest is to study the mechanism 
of $B_{+}$--$B_{-}$ mass splitting.
%We do not use this IF, because we are interested in the effect 
%of the chiral symmetry breaking to the mass splitting of $B_{+}$ 
%and $B_{-}$ and in our method proposed in the Ref.~\cite{jko} 
%$B_{+}$ and $B_{-}$ can be treated at a time.  

%%%%%%%%%%
\section{Determination of the $B_-$ masses}
We have three phenomenological parameters, the mass $m_{B_{\pm}}$, 
the threshold $s_\pm$ and the coupling strength $\lambda_\pm$ to be 
determined from the sum rules (~\ref{eq:srp}) and (~\ref{eq:srn}).
%These parameters are determined by solving the system of three equations, 
%eq. (~\ref{eq:srp}), the first and the second derivatives of 
%(~\ref{eq:srp}) with respect to the Borel mass. 
%%%%%%%%%%%%%%%%%%%%% Added for revised varsion %%%%%%%%%%%%%%%%%%
Unlike the standard sum rule we have only one sum rule for each baryon.
Therefore we are forced to solve the system of three equations, 
eq. (~\ref{eq:srp}), the first and the second derivatives of 
(~\ref{eq:srp}) with respect to the Borel mass.
In general, when we defferntiate the sum rule with respect to 
the Borel mass, the reliability of the QCD sum rule is lost, 
since the derivative picks up the a factor $p^2$, which may enhance 
the contribution of the continuum.
But we should judge the reliability from the stability of the results
against the Borel mass. 
One sees that the baryon masses show enough stability in our analysis.
%%%%%%%%%%%%%%%%%%%%%%%%%%%%%%%%%%%%%%%%%%%%%%%%%%%%%%%%%%%%%%%%%%%

The theoretical side depends on the QCD parameters, such as the quark 
mass and the gauge coupling constant, and also on the other parameters 
that describe the properties of the nonperturbative vacuum of QCD, 
such as the quark and gluon condensates.  We take the chiral limit for 
the up and down quarks, i.e.\ $m_{q}=0$, where we use the symbol $q$ 
for the up and down quarks.
%The acceptable ranges for the parameters ate given at the third column in 
%Table~\ref{tb:qcdpar}. 
We introduce the strange quark mass $m_{s}$, $\chi \equiv 
\vev{\bar{s}s}/\vev{\bar{q}q}$ and $\chi_{5} \equiv \vev{\bar{s}g 
\sigma \cdot G s} / \vev{\bar{q}g \sigma \cdot Gq}$ for the flavor 
SU(3) symmetry breaking.  The gluon condensate is fixed to $\vev{{ 
\alpha_{s} \over \pi} GG} = (0.36 \GeV)^{4}$ since the coefficient is 
small in comparison with the other terms due to a suppression factor of 
$1/(2 \pi)^2$~\cite{lein}.  The vacuum saturation is assumed for 
evaluating the matrix element of the four-quark operators, i.e.\ 
$\vev{(\bar{q}q)^{2}} = \vev{\bar{q}q}^{2}$.  
%These parameters in the 
%theoretical side have some uncertainty, which would come from the 
%truncation of power corrections in the QCD sum rule.  In order to keep 
%consistency in the sum rules, 
%%%%%%%%%%%%%%%%%%%
%Since standard values have some uncertainty,
%we determine these parameters within our sum rules.  
These parameters have some uncertainty, which depends on the truncation 
in the OPE\@. In order to remove this uncertainty, we use our sum rules in
the following way.
%%%%%%%%%%%%%%%%%%%%
The value of $\langle \bar{q} q \rangle$ and $m_{0}^{\;2} 
\equiv \vev{\bar{q} g \sigma \cdot G q}/ \vev{\bar{q}q}$ are 
determined so that the sum rules (~\ref{eq:srp}) and (~\ref{eq:srn}) 
for the nucleon reproduce the observed masses of $N_{+}$ and $N_{-}$.  
In doing so we require that the prediction of the sum rule at $M 
\simeq m_B$ coincides with the observed mass within 5\% and also that 
for the Borel stability variation of the predicted mass against $M$ in 
the region $m_{B} \sim m_{B}+0.5 \GeV$ is less than 10\%.  In the same 
way, the values of $m_{s}$, $\chi$ and $\chi_{5}$ are determined so 
that the sum rules (~\ref{eq:srp}) for the hyperons give the observed 
masses of the $\Lambda_{+}$, $\Sigma_{+}$ and $\Xi_{+}$.

\section{Results and Discussion}
The determined parameters in the theoretical side
are given in Table~\ref{tb:detpar}.
%These parameters are consistent with the standard value. 
The masses of the positive-parity hyperons, $\Lambda_+$, $\Sigma_+$
and $\Xi_+$ are sensitive to the ${\rm SU}_{f}(3)$ breaking parameters 
$m_s$, $\chi$ and $\chi_5$. Therefore these parameters are determined
well. As we shall see later, the value of $\vev{\bar{q}q}$ is determined
from the mass splitting of $B_+$ and $B_-$.
The up and down quark condensate agrees well to the ``standard value'' 
$\vev{\bar{q}q} = (-0.225 \pm 0.025 \GeV)^3$, which was estimated in 
the chiral perturbation~\cite{gl}, and the value $\vev{\bar{u}u} = 
-(0.230 \pm 0.015 \GeV)^3$ that was estimated in the QCD sum rule for 
the octet and the decuplet baryons~\cite{rry2}.  The value $m_{0}$ is 
also consistent with $m_{0}^{\; 2}= 0.5 \sim 1.0 \GeV ^{2}$ estimated 
in the sum rules for baryon~\cite{bi,rry2}, and $m_{0}^{\; 2} = 1.1 \pm 
0.1 \GeV^2$ in the lattice calculation~\cite{ks}.  The instanton 
contribution leads to the mixed condensate somewhat larger, 
$m_{0}^{\;2} = 1.4 \GeV^2$~\cite{pw,ko}.  The reason for this large 
value is that the higher dimensional operators induced by 
the instanton reduce $m_{0}$ effectively~\cite{ko}. 
%%%%%%%%%%%%%%%%%%%%%%% Added for revised version %%%%%%%%%%%%%%%%
Although the strange quark mass $m_s$ is somewhat smaller than the update
analysis~\cite{cps}, our results of the $B_-$ masses are insensitive 
to $m_s$.
%%%%%%%%%%%%%%%%%%%%%%%%%%%%%%%%%%%%%%%%%%%%%%%%%%%%%%%%%%%%%%%%%
The sum rules for the baryons gives $\chi \sim 0.8$~\cite{rry2,bi2}.  
The value of the $\chi_5$ is expected to be close to $\chi$, 
because both $\chi$ and $\chi_5$ are related to the flavor SU(3) 
breaking in QCD\@.  In Ref.~\cite{djn}, however, the sum rule for 
$\Omega$ baryon suggests $\chi_{5} = 1.4$.

\begin{figure}[h]
   \epsfxsize=13cm
   \epsfbox{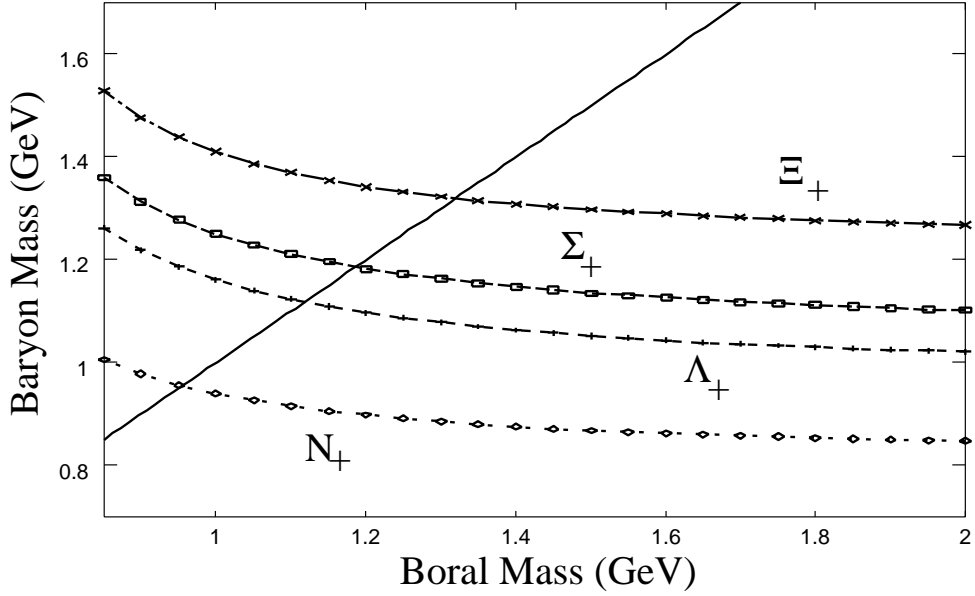}
   \caption[]{
     The Borel mass $M$ dependence of the octet $B_+$ masses.
     The solid line denotes $M = m_{B}$.}
   \label{fig:octp}
\end{figure}

\begin{figure}[h]
   \epsfxsize=13cm
   \epsfbox{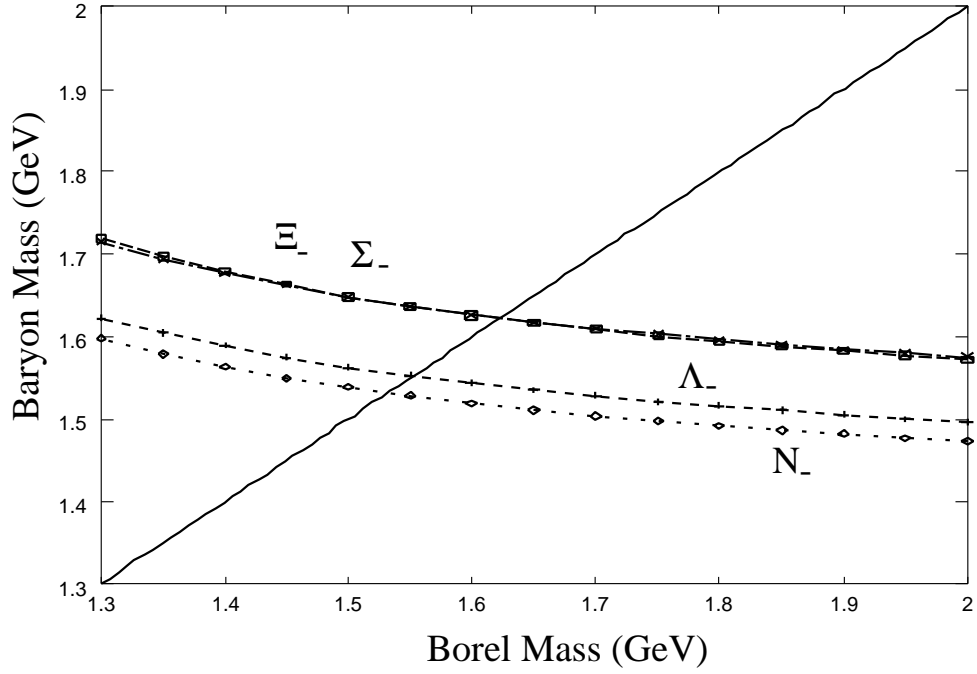}
   \caption[]{
     The Borel mass $M$ dependence of the octet $B_-$ masses.
     The solid line denotes $M = m_{B}$. The lines for 
     masses of $\Sigma_-$ and $\Xi_-$ are almost overlapped.}
   \label{fig:octn}
\end{figure}

\begin{figure}[h]
   \epsfxsize=13cm
   \epsfbox{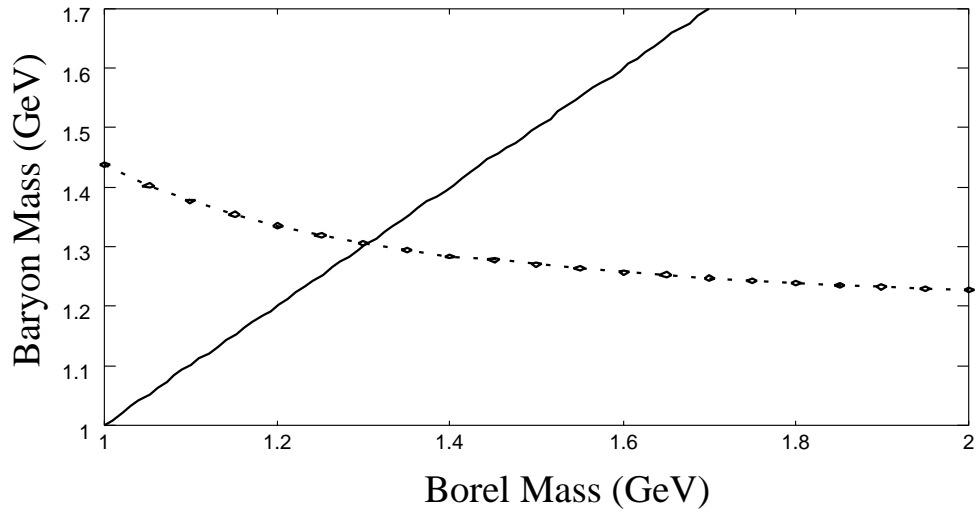}
   \caption[]{
     The Borel mass $M$ dependence of the singlet $\Lambda_S-$ masses.
     The solid line denotes $M = m_{B}$.}
   \label{fig:l0n}
\end{figure}

\begin{figure}[h]
   \epsfxsize=13cm
   \epsfbox{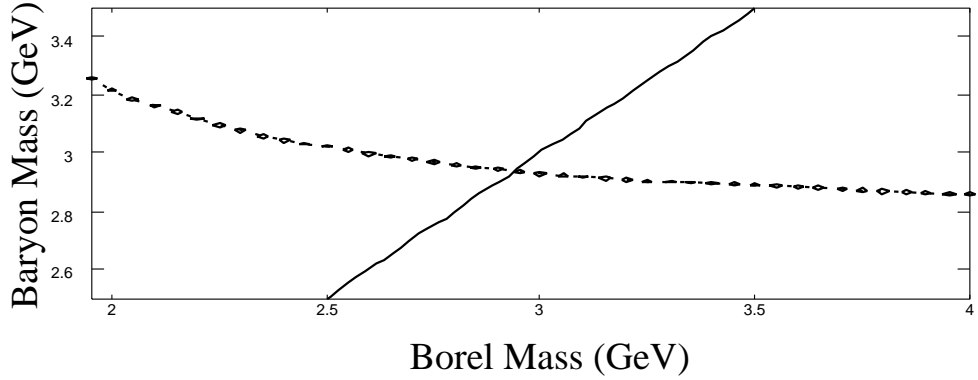}
   \caption[]{
     The Borel mass $M$ dependence of the singlet $\Lambda_S+$ masses.
     The solid line denotes $M = m_{B}$.}
   \label{fig:l0p}
\end{figure}

The masses of the flavor octet and singlet baryons calculated in the 
QCD sum rule are shown in Table~\ref{tb:result}.  The observed masses 
are reproduced fairly well.  The masses of the $\Lambda_-$, 
$\Sigma_-$, $\Xi_-$, $\Lambda_S-$ and $\Lambda_S+$ are the prediction 
without adjustable parameters.  These masses are taken at the Borel 
mass $M \simeq m_B$.  The $M$ dependence of the masses are shown in 
Fig.~\ref{fig:octp} for the octet $B_+$, in Fig.~\ref{fig:octn} for 
the octet $B_-$, in Fig.~\ref{fig:l0n} for $\Lambda_S-$ and in 
Fig.~\ref{fig:l0p} for $\Lambda_S+$.  All masses are stable against 
the Borel mass $M$.  The excited $\Xi$ baryon with $J^{P}= {1 \over 
2}^{-}$ has not been identified by experiment, but resonances with 
unknown spin and parity are found at 1690 MeV and 1950 MeV. The 
prediction of our sum rule prefers $\Xi(1690)$.  Our result suggests 
that the masses of the $B_-$ tend to be degenerate.  This is the 
result of two different origins of the mass difference.  The strange 
quark mass raises the hyperon masses, while
%the quark condensate raises the $B_-$ masses. 
the quark condensate widens the mass splitting of $B_+$ and $B_-$.
Because the strange quark condensate is smaller than the 
up and down quark condensate, the effect of the strange quark mass is partly
canceled in the negative parity baryons.

\begin{figure}[h]
   \epsfxsize=13cm
   \epsfbox{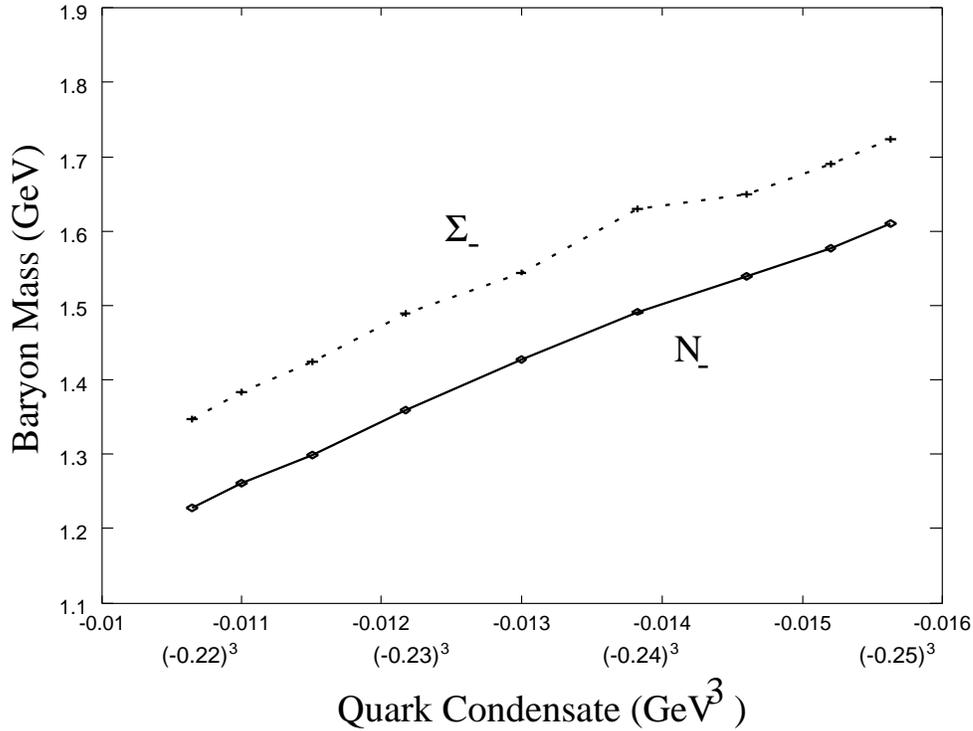}
   \caption[]{The $\vev{\bar{q}q}$ dependence of 
the masses of $N_-$ and $\Sigma_-$.  
The other QCD parameters are fixed so that the masses of the ground 
state baryons $N_+$, $\Lambda_+$, $\Sigma_+$ and $\Xi_+$ are reproduced 
for each value of $\vev{\bar{q}q}$. The value of the mixed 
condensate $\langle \bar{q} g \sigma \cdot G q \rangle = m_{0}^{2} 
\langle \bar{q} q \rangle$ and the strange quark condensate $ 
\langle\bar{s}s\rangle = \chi \langle \bar{q} q \rangle$ are varied 
along with the quark condensate $\vev{\bar{q}q}$.}
   \label{fig:qqdep}
\end{figure}

% We find that the masses of $B_-$ are sensitive to the value of 
% the quark condensate.
The $\vev{\bar{q}q}$ dependence of the masses of $B_{-}$ is shown in 
Fig~\ref{fig:qqdep}, where for each value of $\vev{\bar{q}q}$ the 
other parameters in the theoretical side, $m_0$, $m_s$, $\chi$ and 
$\chi_5$ are adjusted so that the masses of $N_+$, $\Lambda_+$, 
$\Sigma_+$ and $\Xi_+$ are reproduced, while the mixed condensate 
$\langle \bar{q} g \sigma \cdot G q \rangle = m_{0}^{\;2} \langle 
\bar{q} q \rangle$ and the strange quark condensate 
$\langle\bar{s}s\rangle = \chi \langle \bar{q} q \rangle$ are varied 
along with the quark condensate $\vev{\bar{q}q}$.  We see that the 
quark condensate pushes up the masses of the $N_-$ and $\Sigma_-$ as 
long as the masses of $N_+$ and $\Sigma_+$ are fixed.  Therefore the 
magnitude of the quark condensate determines the scale of the mass 
splitting of $B_+$ and $B_-$.  The masses of $\Lambda_{-}$ and 
$\Xi_{-}$ behave similarly against the quark condensate as in $N$ and 
$\Sigma$.

It is extremely interesting to observe that the QCD sum rule predicts 
the flavor-singlet $\Lambda_{S}$ spectrum in the reversed order.  
Namely, the baryon $\Lambda_{S-}$ is lighter than the positive parity 
$\Lambda_{S+}$.  This is consistent with the quark model prediction 
that the Pauli principle forbids all quarks occupying the ground 
s-wave state.  In the correlation function $B$ for the $\Lambda_S$, 
there is no dimension five term, $\langle \bar{q} g \sigma \cdot G q 
\rangle$.  If we put the dimension five term in the $B$ correlation 
function by hand and calculate the masses of the $\Lambda_S+$ and 
$\Lambda_S-$, then we find that the increase of the dimension five 
term raises the mass of $\Lambda_S-$ and lowers that of $\Lambda_S+$.  
Thus we conclude that the absence of the mixed condensate term in the 
$\Lambda_{S}$ sum rule causes the reversed order of $\Lambda_{S+}$ and 
$\Lambda_{S-}$.  We also confirm that the $\vev{\bar{q}g \sigma \cdot 
G q}$ terms, is essential in raising the $B_{-}$ masses as was 
stressed in Ref.~\cite{kimlee}.

\section{Summary}
We have proposed a technique to estimate the masses of negative-parity 
baryon resonances $B_-$ in the QCD sum rule.  It is important to 
separate the $B_-$ contribution from the positive-parity state 
$B_{+}$.  We find that when the chiral symmetry is restored, the 
masses of $B_+$ and $B_-$ become degenerate.  This is quite natural 
from the chiral symmetry point of view and seems to suggest that 
N(1535) (or in general $B_{-}$) is the chiral partner of N(940) (or 
$B_{+}$).  We have calculated the masses of the flavor singlet and 
octet $B_+$ and $B_-$, and have confirmed that the sum rule reproduces 
the observed masses.  The mass of $\Xi$-baryon is predicted 1.63 GeV 
and it may be assigned to the observed $\Xi$(1690) for which the 
spin-parity is not yet known.  We find that the flavor singlet baryon 
with negative parity is the ground state and that the predicted mass 
is close to $\Lambda(1405)$.  We confirm that the magnitude of the 
chiral symmetry breaking vacuum condensate, such as $\vev{\bar{q}q}$ 
and $\vev{\bar{q}g \sigma \cdot G q}$, determines the scale of the 
mass splitting of $B_+$ and $B_-$.

%%%%%%%%%%%%%%%%%%%%% Added for the revised varsion %%%%%%%%%%%%%%%%%%%%%
In the present analysis, we have not calculated the next-to-leading order 
of $\alpha_s$ or higher dimensional terms in the theoretical side.
Although they may modify our numbers slightly, we believe that the qualitative
features of our calculation will not change.  
They are left for future analysis.
%%%%%%%%%%%%%%%%%%%%%%%%%%%%%%%%%%%%%%%%%%%%%%%%%%%%%%%%%%%%%%%%%%%%%%%%

\section*{Appendix}
The ``old fashioned'' correlation function is defined by
\begin{equation} 
   \Pi(p) = i \int \!  d^{4}\!  x e^{ix \cdot p} \theta(x_{0}) 
   \bra{0} J(x) \bar{J}(0) \cket{0}. 
\end{equation}
\begin{equation}
{\rm Im} \Pi(p_{0},\vec{p}=0) \equiv \gamma_{0} A(p_{0}) + B(p_{0}).
    \label{eq:ab}
\end{equation}
The functions $A(p_{0})$ and $B(p_{0})$ defined in 
eq.(~\ref{eq:ab}) up to dimension six and ${\cal O}(m_{s})$ 
neglecting $\alpha_{s}$  are given for each baryon in
the following.
\begin{description}
	\item[{\large N}]  
	\begin{eqnarray}
		A(p_{0}) & = & {5 + 2 t + 5 t^2 \over 2^{10} 
        \pi^{4}} {p_{0}}^{5} \theta (p_{0}) \\
       & &  +{ 5 + 2 t + 5 t^{2} \over 2^{9} \pi^{2}} p_{0} \theta 
(p_{0}) \langle 
        {\alpha_{s} \over \pi} GG \rangle   \nonumber \\
       & & -{5 + 2t -7 t^{2} \over 12 } \delta(p_{0})     
        \langle  \bar{q} q \bar{q} q \rangle , \nonumber \\
		B(p_{0}) & = &  {5 + 2 t - 7 t^{2}\over 32 \pi^{2}} 
{p_{0}}^{2} 
          \theta(p_{0}) \langle \bar{q} q \rangle \\
	   & & -{3(1-t^{2} ) \over 32 \pi^{2}} \theta(p_{0}) \langle \bar{q} g
	  \sigma \cdot G q \rangle . \nonumber
	\end{eqnarray}
	
	\item[\mbox{\large $\Lambda$}]
	\begin{eqnarray}
		A(p_{0}) & = & {5 + 2 t + 5 t^2 \over 2^{10}
        \pi^{4}} {p_{0}}^{5} \theta (p_{0}) \\
	   & &  +{ 5 + 2 t + 5 t^{2} \over 2^{9} \pi^{2}} p_{0} \theta 
	   (p_{0}) \langle {\alpha_{s} \over \pi} GG \rangle  \nonumber 
	   \\
       & & +\left({1 + 4 t - 5 t^{2} \over 48 \pi^{2}} m_{s} \langle 
\bar{q}q 
        \rangle + {5+2t+5t^{2} \over 64 \pi^{2}} m_{s}\langle 
        \bar{s} s \rangle \right) p_{0}\theta(p_{0}) \nonumber \\
		& &  +\left(-{13-2t-11 t^{2} \over 36} \langle \bar{q}q\bar{q}q 
		\rangle - {1 + 4 t - 5 t^{2} \over 18} \langle 
		\bar{s}s\bar{q}q \rangle \right) \delta(p_{0})  \nonumber \\
	   & & +\left(-{5 +2t -7 t^{2} \over 96 \pi^{2}} \delta (p_{0}) + 
	   {1-t^{2} \over 32 \pi^{2}} {\theta ( p_{0}^{2} - m_{s}^{2}) 
	   \over p_{0}} \right) m_{s} \langle \bar{q} g \sigma \cdot G 
	   q\rangle \nonumber \\
	   & & - {1+t+t^{2} \over 48 \pi^{2}} m_{s}\langle \bar{s} g 
	   \sigma \cdot G s\rangle \delta(p_{0}) \nonumber \\
		B(p_{0}) & = & - {13 - 2t -11 t^{2} \over 3 \cdot 2^{8} 
\pi^{4}}m_{s} 
		{p_{0}}^{4} \theta(p_{0}) \\
		& & +\left({1 + 4 t - 5t^{2} \over 48 \pi^{2}} 
		\langle\bar{q}q\rangle +
		{13-2t -11 t^{2}\over 96 \pi^{2}} \langle \bar{s}s \rangle \right) 
{p_{0}}^{2} 
		\theta(p_{0})   \nonumber \\
		& &-{1 - t^{2} \over 32 \pi^{2}} ( \langle \bar{q} g
	  \sigma \cdot G q \rangle + 2 \langle \bar{s} g
	  \sigma \cdot G s \rangle )  \theta(p_{0}). \nonumber
	\end{eqnarray}

%	\pagebreak[4]

	\item[\mbox{\large $\Sigma$}]  
	\begin{eqnarray}
		A(p_{0}) & = & {5 + 2 t + 5 t^2 \over 2^{10}
        \pi^{4}} {p_{0}}^{5} \theta (p_{0}) \\
       & &  +{ 5 + 2 t + 5 t^{2} \over 2^{9} \pi^{2}} p_{0} \theta 
(p_{0}) \langle 
        {\alpha_{s} \over \pi} GG \rangle   \nonumber \\
       & & +\left({3 (1 - t^{2}) \over 16 \pi^{2}} m_{s} \langle 
\bar{q}q 
        \rangle + {5+2t+5t^{2} \over 64 \pi^{2}} m_{s}\langle 
        \bar{s} s \rangle \right) p_{0}\theta(p_{0}) \nonumber \\
       & &  +\left({1-2t+t^{2} \over 12} \langle \bar{q}q\bar{q}q 
\rangle - 
        {1-t^{2} \over 2} \langle \bar{s}s\bar{q}q \rangle \right) 
\delta(p_{0}) \nonumber \\
	   & &  +{1-t^{2}\over 32 \pi^{2}} \left(- 8  \delta (p_{0}) + {3 
	   \over p_{0}} \theta ( p_{0}^{2} - m_{s}^{2}) 
	   \right) m_{s} \langle \bar{q} g \sigma \cdot G q\rangle  
	   \nonumber \\
	   & &+{1+t+t^{2} \over 48 \pi^{2}} m_{s}\langle \bar{s} g \sigma 
	   \cdot G s\rangle \delta(p_{0}) \nonumber \\
		B(p_{0}) & = & {1-2t+t^{2} \over 2^{8} \pi^{4}}m_{s} 
		{p_{0}}^{4} \theta(p_{0}) \\
		& & +\left({3(1-t^{2}) \over 16 \pi^{2}} \langle\bar{q}q\rangle - 
		{1-2t+t^{2}\over 32 \pi^{2}} \langle \bar{s}s \rangle \right) 
{p_{0}}^{2} 
		\theta(p_{0})  \nonumber  \\
		& &-{3(1 - t^{2}) \over 32 \pi^{2}} \theta(p_{0}) \langle \bar{q} g
	  \sigma \cdot G q \rangle . \nonumber
	\end{eqnarray}

	\item[\mbox{\large $\Xi$}]  
	\begin{eqnarray}
		A(p_{0}) & = & {5 + 2 t + 5 t^2 \over 2^{10}
        \pi^{4}} {p_{0}}^{5} \theta (p_{0})  \\
        & &+{ 5 + 2 t + 5 t^{2} \over 2^{9} \pi^{2}} p_{0} \theta 
(p_{0}) \langle 
        {\alpha_{s} \over \pi} GG \rangle \nonumber \\
        & &  +\left({3 (1 - t^{2}) \over 16 \pi^{2}} m_{s} \langle 
\bar{q}q 
        \rangle + {3 (1+2t+t^{2}) \over 32 \pi^{2}} m_{s}\langle 
        \bar{s} s \rangle \right) p_{0}\theta(p_{0}) \nonumber \\
		 &  &  +\left({1-2t+t^{2} \over 12} \langle \bar{s}s\bar{s}s \rangle 
- 
        {1-t^{2} \over 2} \langle \bar{s}s\bar{q}q \rangle \right)  
\delta(p_{0}) \nonumber \\
        & & +{1-t^{2} \over 32 \pi^{2}}\left(- 4\delta(p_{0}) + {3 
        \over p_{0}} \theta(p_{0}^{2} - 4 m_{s}^{2}) \right) m_{s} 
        \langle \bar{q} g \sigma \cdot G q\rangle  \nonumber \\
        & & -{1+10t+t^{2} \over 192 \pi^{2}}  
        m_{s}\langle \bar{s} g \sigma \cdot G s\rangle 
         \delta(p_{0}) \nonumber \\
		B(p_{0}) & = & {-3(1-t^{2}) \over 2^{7} \pi^{4}}m_{s} 
		{p_{0}}^{4} \theta(p_{0})  \\
		& & +\left({3(1-t^{2}) \over 16 \pi^{2}} \langle\bar{s}s\rangle - 
		{1-2t+t^{2}\over 32 \pi^{2}} \langle \bar{q}q \rangle \right) 
{p_{0}}^{2} 
		\theta(p_{0}) \nonumber \\
		& & -{3(1 - t^{2} ) \over 32 \pi^{2}} \theta(p_{0}) \langle \bar{s} g
	  \sigma \cdot G s \rangle . \nonumber
	\end{eqnarray}
	
	\item[\mbox{\large $\Lambda_{S}$} (Flavor Singlet)]
	\begin{eqnarray}
		A(p_{0}) & = & {3 \over 2^{8}
        \pi^{4}} {p_{0}}^{5} \theta (p_{0}) \\
	   & &  +{ 3 \over 2^{7} \pi^{2}} p_{0} \theta 
	   (p_{0}) \langle {\alpha_{s} \over \pi} GG \rangle \nonumber 
	   \\
       & & + {1 \over 16 \pi^{2}} m_{s} (4 \langle \bar{q}q 
        \rangle + 3 \langle \bar{s} s \rangle ) 
        p_{0}\theta(p_{0})  \nonumber \\
		& & -{1 \over 3} ( \langle \bar{q}q\bar{q}q 
		\rangle + 2 \langle \bar{s}s\bar{q}q \rangle) \delta(p_{0}) 
\nonumber \\
	   & &  -{1 \over 16 \pi^{2}} m_{s} (\langle \bar{q} g \sigma \cdot G 
	   q\rangle + 3 \langle \bar{s} g \sigma \cdot G 
	   s\rangle) \delta (p_{0}) \nonumber \\
		B(p_{0}) & = & - {1 \over 64 \pi^{4}}m_{s} 
		{p_{0}}^{4} \theta(p_{0}) \\
		& &  +{1\over 8 \pi^{2}} (
		2 \langle\bar{q}q\rangle + \langle \bar{s}s \rangle ) {p_{0}}^{2} 
		\theta(p_{0}) \nonumber
	\end{eqnarray}

\end{description}

%\begin{table}[h]
%	\caption{The QCD parameters in sum rules}
%	\begin{tabular}{llll}
%\hline \hline
% quark condensate &  $ \langle \bar{q} q \rangle  $
%  & $-(0.20 \sim 0.25 \GeV)^{3}$ & $-(0.225 \pm 0.025 \GeV)^3$\cite{gl} \\
%  & & & $-(0.22 \sim 0.25 \GeV)^{3}\cite{ept}$ \\
%  &  $\chi \equiv \frac{\langle \bar{s} s \rangle}{\langle \bar{q} q 
%\rangle}$ & $0.5 \sim 1.0$ &   \\
% mixed condensate & $ m_{0}^{\; 2} \equiv 
%     \frac{ \langle \bar{q}g\sigma\cdot G q \rangle}{ \langle\bar{q} q 
%\rangle}$ & $(0.7 \sim 1.0 \GeV)^2$  & \\
%  & $ \chi_{5} \equiv \frac{\langle \bar{s} g\sigma \cdot G s \rangle}{
%     \langle \bar{q} g \sigma \cdot G q \rangle}$
%	& $0.5 \sim 1.4 $ & \\
% quark mass & $ m_{u} , m_{d}$ 
%	& 0  & \\
%  & $ m_{s} $ 
%	& $0.1 \sim 0.25 \GeV$  & \\
% gluon condensate & $ \langle {\alpha_{s} \over \pi} GG \rangle $ 
%  & $(0.36 \GeV)^{4}$  (fixed) &  \\
%\hline \hline
%\end{tabular}
%	\protect\label{tb:qcdpar}
%\end{table}

\begin{table}[h]
  \caption{The determined QCD parameters}
  \begin{center}
  \begin{tabular}{ccccc}
	\hline \hline
	$\langle\bar{q}q\rangle$ & $m_{0}$ & $m_{s}$ & $\chi$ & $\chi_{5}$  \\

	(-0.244 GeV$)^{3}$ & 0.9 GeV & 0.1 GeV & 0.75 & 0.8   \\
	\hline \hline
	\end{tabular}
  \end{center}
 	\protect\label{tb:detpar}
\end{table}

\begin{table}[t]
\begin{center}
  \caption{}
  \begin{tabular}{r|cccccccc|cc}
    \multicolumn{11}{r}{Unit: GeV} \\
	 \hline	\hline
    Baryon & $N_+$ &  $\Lambda_+$ & $\Sigma_+$ & $\Xi_+$ & 
            $N_-$ & $\Lambda_-$ & $\Sigma_-$ & $\Xi_-$ 
          & $\Lambda_{S-}$ & $\Lambda_{S+}$	\\
	Sum rule & 0.94 & 1.12 & 1.21 & 1.32 & 
	           1.54 & 1.55 & 1.63 & 1.63 & 1.31 & 2.94 \\
	Exp. & 0.94 & 1.12 &1.19  & 1.32 &
	          1.535 & 1.67 & 1.62 & -----& 1.405 & ----- \\
	\hline \hline            
  \end{tabular}
	  \label{tb:result}
\end{center}
\end{table}

\end{document}